# CURRENT NOISE INVESTIGATIONS IN JOSEPHSON DEVICES BY SWITCHING CURRENT MEASUREMENTS

C. Granata[a], A. Vettoliere[b], R. Russo, M. Russo, and B. Ruggiero

*Istituto di Cibernetica "E. Caianiello" del Consiglio Nazionale delle Ricerche, I-80078 Pozzuoli (Napoli), Italy*

**Abtract:**

An experimental investigation of the critical current noise in underdamped niobium based Josephson junctions by a technique based on the switching current measurements is reported. By swiping the junction with a current ramp we measure the critical current switching using the standard time of flight technique. In such a way it is possible to obtain a digital sampling of the critical current as a function of the time $I_c(t)$ and the corresponding fluctuation $\Delta I_c = I_c(t) - \langle I_c(t) \rangle$ with a sampling frequency given by the ramp frequency. Computing the Fast Fourier Transform module it is possible to evaluate the power spectral density of the critical current fluctuation $S_{\Delta I_c}(f)$. Measurements on Josephson junctions having an area ranging from $(4 \times 4)$ µm$^2$ to $(40 \times 40)$ µm$^2$ in the temperature range from 4.2 K to 1.2 K are reported. The experimental results show a linear behavior of the current white noise from both the junction area and the temperature. A very good agreement between the experimental data and the Nyquist noise of the junction normal resistance is shown. These measurement provide very useful information about the intrinsic noise of Josephson devices involving SQUIDs and qubits.

**Pacs: 74.50+r, 85.25.Dq, 85.25.Cp, 74.78.Na**

One of the most important figure of merit of a Josephson device is the noise which is typically expressed by the spectral density of the equivalent current, voltage, magnetic flux or field, or charge noise. Due to the extremely intrinsic low noise, the Josephson devices (in particular Superconducting QUantum Interference Devices - SQUIDs) have been employed in many applications like, biomagnetism, metrology, magnetic microscopy, astrophysics, nanomagnetism, quantum computing, and particle Physics [1, 2]. The importance of the noise in Josephson devices has stimulated many theoretical and experimental investigations leading to an exhaustive comprehension of the main mechanisms responsible of the different noise [3]. Theories for voltage, current and magnetic flux noise in resistively shunted junctions, rf-SQUID and dc SQUID have been successfully developed [3] as well as quantum charge noise in cooper pair box [4]. In the recent years many theoretical [5-8] and experimental [9-12] investigations have been devoted to the understanding the low frequency noise in Josephson device. Such a kind of noise is a very interesting issue since it is undoubtedly related to the decoherence time in Josephson qubits [13]. However, while there are many experimental investigations of critical current noise in overdamped Josephson junctions, it is not the same for the underdamped case [9], despite both phase and flux Josephson qubits employ the underdamped junctions [14]. Furthermore, even if, it is reasonable to assume that the origin of the white critical current noise in underdamped Josephson junction is the same of the overdamped junction (shunt resistor fluctuations), unambiguous direct measurements of white critical current noise are not available and the involved resistor is not identified.

In this letter, an experimental investigation of the critical current noise in underdamped niobium based Josephson junctions by a technique based on the switching current measurements is presented. A digital sampling of the critical current time oscillation is obtained by swiping the junction with a current ramp out of the superconducting state and measuring the critical current using the standard time of flight technique. In such a way the time critical current function ($I_c(t)$) is digitally sampled with a sampling frequency $f_s$ given by the ramp frequency. If $\Delta t$ is acquisition time, the total sample number is $N=f_s \cdot \Delta t$. The current fluctuation is $\Delta I_c^k = I_c^k - <I_c>$ where k is an index which varies from 0 to N-1 and $<I_c> = (I_0+I_1+...I_{N-1})/N$ is mean value of the critical current. The power spectral density (PSD) of the critical current fluctuation $S_{\Delta Ic}$ (f) is given by the module of the Fast Fourier Transform $X_q$ divided the the total number of samples:

$$S_{\Delta I_c}(q) = \frac{X_q \cdot X_q^*}{N} \qquad X_q = \sum_{k=0}^{N-1} \Delta I_c^k e^{-j\frac{2\pi}{N}kq} \qquad q = 0, 1, ..., N-1 \tag{1}$$

The computation of the PSD is based on the Welch method and the physical bandwidth of the system (junction plasma frequency in our case) has been taken into account.

The maximum frequency content $f_M$ is given by half of the frequency sampling ($f_M=f_s/2$ - Shannon-Nyquist theorem), while the minimum frequency $f_m$ depends from the acquisition time $\Delta t$ ($f_m=1/\Delta t$). The frequency resolution corresponds to the minimum frequency $\Delta f=1/\Delta t$. With respect to the other technique where the underdamped junction (biased above the gap voltage) is placed as one arm of a Wheastone bridge having a SQUID as a null detector [9], this technique allows a direct measure of the critical current oscillations.

The analysis has been performed on high quality underdamped Josephson junctions having an area ranging from (4x4) $\mu m^2$ to (40x40) $\mu m^2$ in the temperature range from 4.2 K to 1.2 K. The sample fabrication process has been well described elsewhere [15], so only an outline of the main fabrication steps is reported here. The Nb/Al-AlOx/Nb trilayer is deposited by dc-magnetron sputtering onto a 3" oxidized Si wafers (380 $\mu$m thick + 300 nm thermal SiO$_2$) in a UHV thin film deposition station. The base and the top niobium electrodes have a thickness of 200 nm and 35 nm, respectively and the Al layer has a thickness of about 7 nm. The artificial (AlOx) tunnel barrier is obtained by a thermal oxidation using dry oxygen. The junction geometry and the relative insulation are obtained by a standard photolithography and a Selective Niobium Anodization Process (SNAP) respectively. In order to improve the junction quality, a further insulation is provided by a SiO$_2$ film and patterned by a lift-off process. Finally, a Nb film (500 nm thick), provides for the junction wiring and contact pads. This procedure is capable to routinely produce high quality window-type junctions having an area ranging from (3x3) µm² to (100x100) µm².

The experimental set up is shown in Fig. 1. The junction was biased with a triangular-shaped waveform at a frequency of 100 Hz. The synchronism of the ramp generator was delayed and sent to start input of a time acquisition board having a time resolution of 12 ns. The junction voltage was amplified and sent to a discriminator that provides the stop signal at the time of the switching out of the zero voltage state [16]. The critical current values are obtained multiplying the current ramp slope dI/dt (measured after each measurement) with the interval time Δt measured by the time acquisition board. The estimated measurement resolution of the critical current is about 1 part in $10^4$, which is essentially limited by the stability of the synchronism signal and of the delay. The measurements were performed, in a pumped liquid $^4$He cryostat with two copper and three µ-metal coaxial shielding cans, using a low-noise 4-contact current-voltage technique. The sample was mounted on a chip carrier, which contained an integrated high frequency filtering stage located very close to the junction. All the electrical connections to room temperature went through manganine wires.

In the figure 2, a current-voltage characteristic of a (20x20) µm² Josephson junction measured at T=4.2 K, is reported. The inset shows the critical current oscillation measured with the technique described above by using a biasing waveform frequency of 100 Hz. Each measurement includes 100000 samples corresponding to an acquisition time of 1000 *s*. For clarity only a time window of a 1 *s* is reported in the inset of Fig.2, so that it is possible to see the single sampling points. In order to verify the low noise of measurement set-up, the current distributions, obtained by making the histograms of the $\Delta I_c^k$, have been compared with the predictions of the thermal regime theory [16]. The good agreement between the experimental data and the theory guarantees the reliability of the measurements.

In figure 3, the current noise spectra of three junction having an area of (4x4) $\mu m^2$, (20x20) $\mu m^2$ and (40x40) $\mu m^2$ are reported. They have been obtained by averaging 30 different measurements at T=4.2 K for each junction. An acquisition time of 1000 s has allowed us to reach a frequency as low as a 1 mHz with a resolution of a 1 mHz. For each junction, it is possible to observe a white noise and a frequency dependent noise with a knee depending on the junction size. The white current noise $S_{Ic}$ ranges from $1.8 \times 10^{-24}$ $A^2$/Hz for the smallest junction to $1.4 \times 10^{-22} A^2$/Hz for the greatest one. Such values are consistent with the data recently reported in the literature [9]. As expected, the low frequency noise exhibits a 1/f behaviour and an inversely proportionality with the junction area. However the knee position at very low frequency with respect to the data reported in the literature is not clear and it could be due to the employed technique. Note that, as for the recent measurements [9], also in our case the current noises $\sqrt{S_{I_c}(1Hz)}$ of all investigated junctions measured at T=4.2 K, is about one order of magnitude less than the value predicted by the empirical formula [12 $(I_C/\mu A)/(A^{1/2}/\mu m)$ (pA)/Hz$^{1/2}$] obtained by averaging over a wide range of junction areas and critical currents for several different technologies [13].

In the figure 4, the white current noise as a function of junction area with a linear fit is reported. It is clear from the figure that the noise scales with the junction area. Since it is reasonable that the white oscillation of the critical current are due to the resistor fluctuations and considering that the normal resistance of the junction shows a clear linear dependence on the junction area, we have reported in the inset of the Fig. 4 the same noise behavior normalized to the Nyquist noise $S_I(R_N)=4k_BT/R_N$ where $k_B$ is the Boltzman constant and $R_N$ is the normal resistance of the corresponding junction. It is possible to observe from the figure that the agreement is very good, indicating that the white current noise in the Josephson junctions is due to the Nyquist noise of the $R_N$. As a further confirmation, a temperature dependence of the white current noise of a 16 $\mu m^2$ junction has been measured (Figure 5). The experimental data have been fitted with a linear curve $4k_BT/R$. The best fit has yield a value of R= 130 Ω which is close to the normal resistance value ($R_N$=110 Ω). In order to achieve a straight check of our interpretation, a 100 μm$^2$ junction having a $R_N$=20 Ω has been shunted with a external resistor $R_e$<$R_N$, and the noise spectra measured before and after the shunting operation have been compared. After the shunting the junction showed an hysteresis due to a $β_c$ value still greater than 1 and a $R_N$ value about four time smaller than the initial value. As expected an increase of noise of about a factor 4 has been observed in the shunted junction, guaranteeing the reliability of our description. It is worth to note that, if we consider the limit case of the shunted junction without hysteresis ($β_c$<1), in the thermal regime ($eV<<k_BT$, $e$ is the electron charge and $V$ is the junction voltage), the white current noise including the mixed down effect is $S_I=S_V/R_d^2=(1+ I_c/2I)(4k_BT/R_S)$ where $R_d$ is the dynamical resistance, $R_S$ is the shunt resistor and I is bias current [17]. Neglecting the mixed down term (I>>$I_c$), the junction noise coincides with the Nyquist noise of the shunt resistor. In analogy with the shunted junction, for very low temperature ($eV>>k_BT$), we expect that the quantum effect becomes dominant and the current noise saturates at value limited solely by zero-point fluctuations in the $R_N$ [17].

In conclusion, direct measurements of critical current noise in extremely underdamped niobium based Josephson junctions have been performed. The technique employed is based on the digitally sampling of the critical current by switching current measurements and the computation of the power spectral density. The experimental data undoubtedly indicate that the white noise is due essentially to the Nyquist noise of the junction normal resistance. The low frequency noises scales with the junction area and exhibit a 1/f behavior. These investigations are very useful for Josephson devices including underdamped junction like SQUID triggers, phase or flux qubits and recent interesting applications employing underdamped Josephson junction as a detector of current noise [18].

**Acknowledgements**

This work was partially supported by Italian MiUR under the project "Sviluppo di componentistica superconduttrice avanzata e sua applicazione a strumentazione biomedica" (L 488/92, Cluster 14 - Componentistica Avanzata).

**Figure Captions**

Figure 1

Experimental set up for the switching current measurements based on a time of flight technique. The resolution of the critical current measurements is about 1 part in $10^4$.

Figure 2

Current-voltage characteristic of a underdamped Josephson junction having an area of (20x20) $\mu m^2$ measured a T=4.2 K. The inset shows the values of critical current as function of the time obtained by biasing the junction with a triangular waveform at 100 Hz.

Figure 3

Spectral density of the critical currents relative to three underdamped Josephson junction measured at T=4.2 K. The junction areas are: 16 $\mu m^2$ (green/lower curve), 400 $\mu m^2$ (blue/middle curve) and 1600 $\mu m^2$ (red/higher curve). Each curve has been obtained by averaging the spectral density of 30 different measurements.

Figure 4

White critical current noise measured at T=4.2 K as a function of the Josephson junction area with a linear fit. The inset shows the same data normalized to the Nyquist noise of the corresponding junction normal resistance.

Figure 5

White critical current noise of a (4x4) $\mu m^2$ underdamped Josephson junction as a function of the temperature. The straight line refers to a linear best fit ($S_I=4k_BT/R$) crossing the origin; from the curve slope a value of R=130 $\Omega$ is obtained.

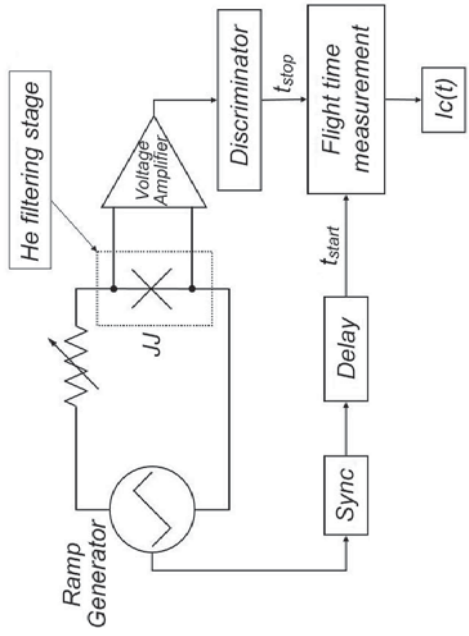

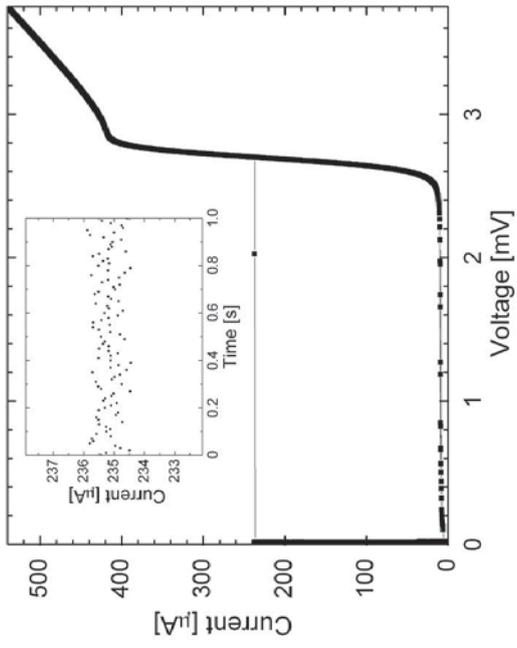

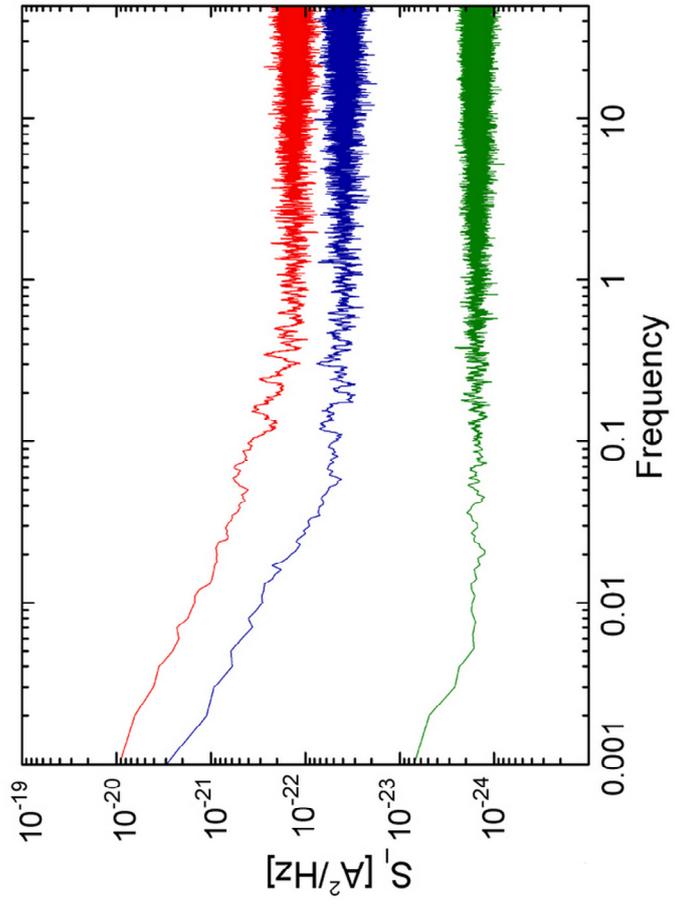

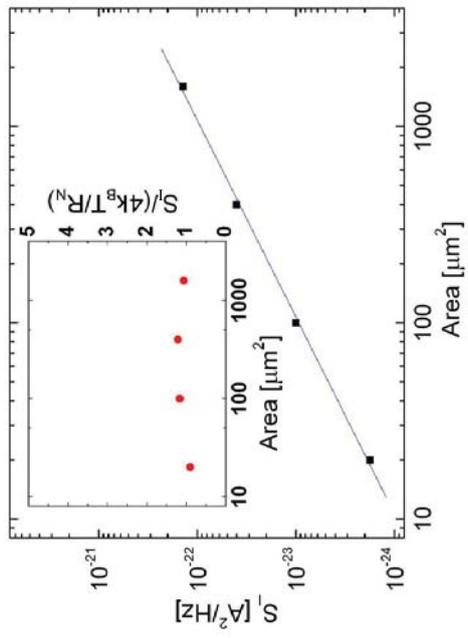

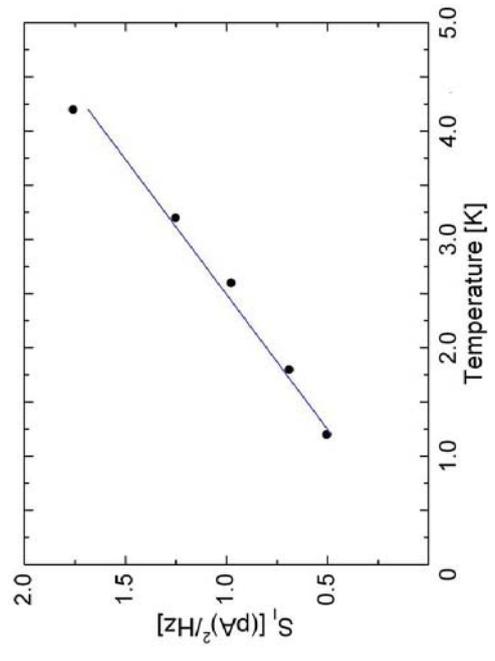